\documentclass[reprint,amsmath,amssymb,aps,floatfix,groupedaddress,superscriptaddress,nofootinbib,preprintnumbers]{revtex4-1}
\usepackage{float}
\usepackage{amsmath}
\usepackage{graphicx,color}
\usepackage{dcolumn}
\usepackage{bm}
\usepackage{here}
\usepackage{comment}
\usepackage[paperwidth=210mm,paperheight=297mm,centering,hmargin=1.5cm,vmargin=2.0cm]{geometry}
\usepackage{tabularx}

\allowdisplaybreaks[1]

\makeatletter
\newsavebox{\@brx}
\newcommand{\llangle}[1][]{\savebox{\@brx}{\(\m@th{#1\langle}\)}%
  \mathopen{\copy\@brx\mkern2mu\kern-0.9\wd\@brx\usebox{\@brx}}}
\newcommand{\rrangle}[1][]{\savebox{\@brx}{\(\m@th{#1\rangle}\)}%
  \mathclose{\copy\@brx\mkern2mu\kern-0.9\wd\@brx\usebox{\@brx}}}
\makeatother

\begin{document}

\title{ Reconstructing particle number distributions with convoluting volume fluctuations } 

\author{ShinIchi Esumi}
\email{esumi.shinichi.gn@u.tsukuba.ac.jp}
\affiliation{Tomonaga\,Center\,for\,the\,History\,of\,the\,Universe,\,University\,of\,Tsukuba,\,Tsukuba,\,Ibaraki\,305,\,Japan}
\author{Kana Nakagawa}
\affiliation{Tomonaga\,Center\,for\,the\,History\,of\,the\,Universe,\,University\,of\,Tsukuba,\,Tsukuba,\,Ibaraki\,305,\,Japan}
\author{Toshihiro Nonaka}
\email{tnonaka@rcf.rhic.bnl.gov}
\affiliation{Tomonaga\,Center\,for\,the\,History\,of\,the\,Universe,\,University\,of\,Tsukuba,\,Tsukuba,\,Ibaraki\,305,\,Japan}
\affiliation{Key\,Laboratory\,of\,Quark\,\&\,Lepton\,Physics\,(MOE)\,and\,Institute\,of\,Particle\,Physics,\,Central\,China\,Normal\,University,\,Wuhan\,430079,\,China}


\begin{abstract}
We propose methods to reconstruct particle distributions 
with and without considering initial volume fluctuations. 
This approach enables us to correct for detector efficiencies and initial volume 
fluctuations simultaneously. 
Our study suggests such a tool could investigate the possible bimodal structure of net-proton distribution 
in Au+Au collisions at $\sqrt{s_{\rm NN}}=$~7.7~GeV as a signature of 
first-order phase transition and critical point of hadronic matter~\cite{Bzdak:2018uhv,Bzdak:2018axe}.
\end{abstract}
\maketitle

\newcommand{\ave}[1]{\ensuremath{\langle#1\rangle} }

\onecolumngrid

\section{Introduction}
In recent years, the higher-order cumulants of net-particle distributions 
have been actively measured in heavy-ion collision experiments to study 
QCD phase structure~\cite{bookKoch,Asakawa:2015ybt,Luo:2017faz,Bzdak:2019pkr}.
Higher-order cumulants ($C_{r}$, $r\leq6$) and their ratios of net-particle distributions 
were reported by ALICE~\cite{Arslandok:2020mda}, HADES~\cite{Adamczewski-Musch:2020slf}, 
NA61/SHINE~\cite{Mackowiak-Pawlowska:2020glz} and STAR experiments~\cite{Aggarwal:2010wy,net_proton,net_charge,Adamczyk:2017wsl,Nonaka:2020crv,Adam:2020unf}.
In particular, the non-monotonic beam energy dependence of net-proton $C_{4}/C_{2}$ 
observed in Au+Au central collisions at the STAR experiment~\cite{Adam:2020unf}
could indicate a possible signature of the critical point 
in low collision energies.

Recently, it was pointed out that 
the strong enhancement of $C_{4}/C_{2}$ 
of the net-proton distributions observed 
in Au+Au central collisions at $\sqrt{s_{\rm NN}}=$~7.7~GeV by the STAR 
experiment~\cite{Adam:2020unf} 
can be explained by the superposition of binomial and Poisson 
distributions~\cite{Bzdak:2018uhv,Bzdak:2018axe}. Let us refer to this as a "bimodal" distribution throughout this paper. 
In Ref.~\cite{Adam:2020unf}, cumulants are corrected for detector efficiencies 
analytically by assuming that they follow binomial distributions
~\cite{eff_kitazawa,eff_koch,eff_psd_volker,eff_xiaofeng,Nonaka:2017kko}.
As this approach does not correct the net-proton distribution itself, 
a so-called unfolding approach is necessary to reconstruct the distribution
in terms of detector efficiencies numerically and check the bimodal prediction.

On the other hand, there is a need to also consider other effects such as initial volume fluctuaions.
Various studies have been carried out to understand and correct for 
initial volume fluctuations~\cite{Luo:2013bmi,Gorenstein:2011vq,Skokov:2012ds,Braun-Munzinger:2016yjz,Sombun:2017bxi,Rogly:2018ddx,Broniowski:2017tjq,Sugiura:2019toh}, but there has been no method to remove the volume fluctuations from the distributions.

The purpose of the present study is to present a new method to reconstruct the distribution 
by addressing both volume fluctuations and detector inefficiencies.
Throughout this paper, we consider 
the number of generated particles and antiparticles, 
and measured particles after passing through the detectors, 
denoted by ($N_{p}$,$N_{{\bar p}}$) and ($n_{p}$,$n_{{\bar p}}$), respectively. 
The relation of these variables is given by
\begin{equation}
	P(N_{p},N_{\bar p}) = \sum_{n_{p},n_{{\bar p}}}{\cal R}_{\rm rev}(N_{p},N_{{\bar p}};n_{p},n_{{\bar p}}){\tilde P({n_{p},n_{{\bar p}}})},
\end{equation}
where $P(N_{p},N_{\bar p})$ and $\tilde P(n_{p},n_{\bar p})$ are two-dimensional probability distribution functions 
and ${\cal R}_{\rm rev}(N_{p},N_{{\bar p}};n_{p},n_{{\bar p}})$ is the 
conversion matrix from the measured to generated coordinates, which we call 
the "reversed response matrix" for the rest of this paper. 

We also use the term "detector filter", which represents the Monte-Carlo way to 
determine ($n_{p}$,$n_{{\bar p}}$) from a given ($N_{p}$,$N_{{\bar p}}$).
In this paper, the detector filter has efficiencies following the binomial distribution for simplicity: 
\begin{equation}
	B(n;\varepsilon,N) = \varepsilon^{n}(1-\varepsilon)^{N-n}\frac{N!}{n!(N-n)!},
\end{equation}
where $n$ and $N$ are measured and generated particles, and $\varepsilon$ is the detector efficiency.
Efficiencies for particles and antiparticles will be denoted separately by $\varepsilon_{p}$ 
and $\varepsilon_{\bar p}$, and are assumed to be independent. 
Note that any efficiency including non-binomial distribution~\cite{binomial_breaking,Nonaka:2018mgw} can be properly taken into account in our 
unfolding approach as will be briefly discussed in the following section.

The paper is organized as follows. 
In Sec.~\ref{sec:sec2}, we introduce procedures to reconstruct 
the particle and antiparticle number distributions in terms of the detector efficiencies. 
The method is demonstrated in toy models for both extreme and realistic cases.
In Sec.~\ref{sec:sec3}, we discuss how to implement and correct 
for initial volume fluctuations as well as detector efficiencies.

\section{Particle number unfolding\label{sec:sec2}}
In this section, we demonstrate the approach of particle number unfolding 
through Monte-Carlo simulations, followed by brief discussions on 
the smoothing and scaling of the response matrices.
\subsection{Methodology\label{sec:critical}}
\begin{figure*}[htbp]
	\begin{center}
	\includegraphics[width=160mm]{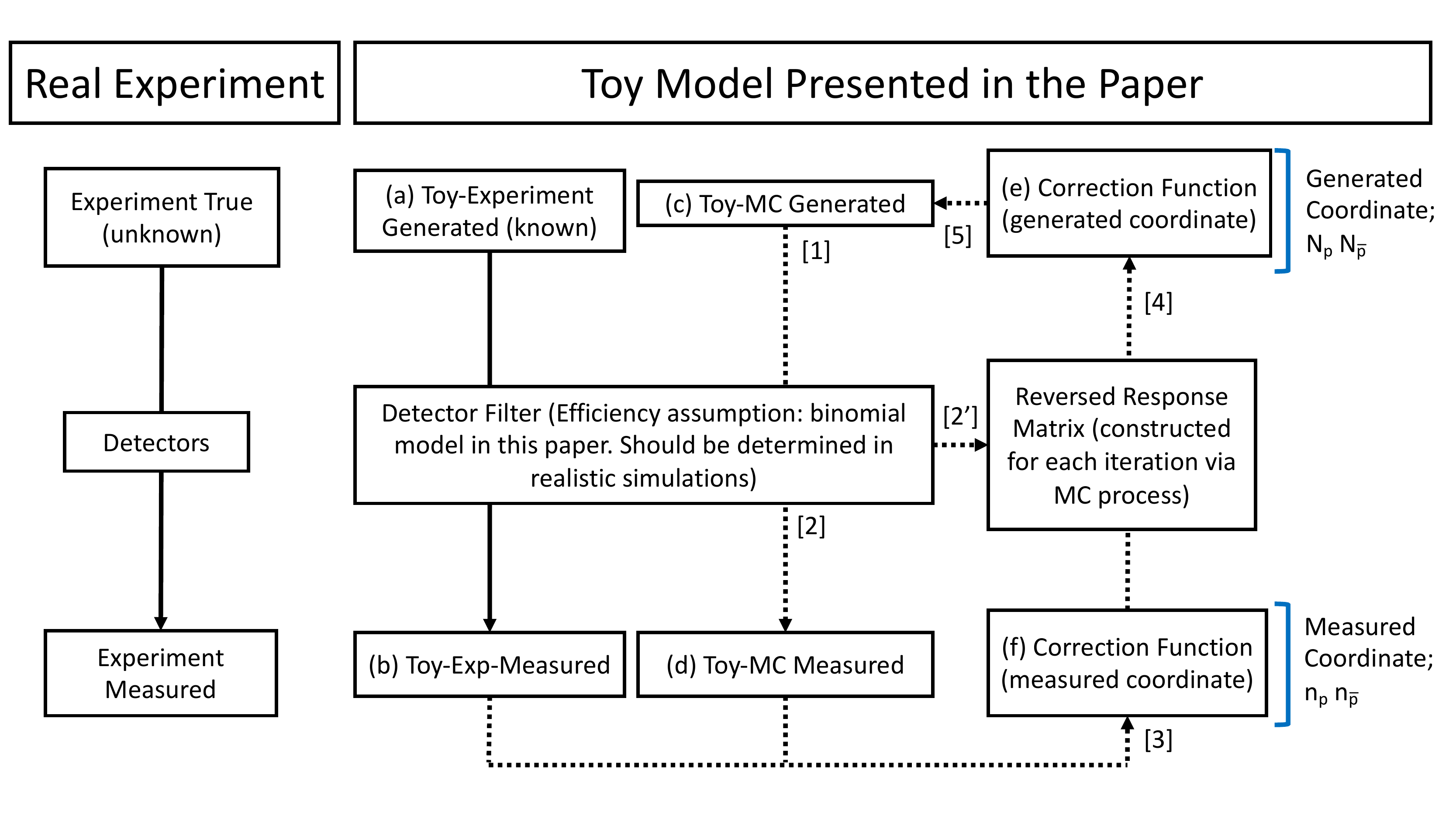}
	\end{center}
	\caption{
		Flowcharts of the unfolding approach. The left hand side corresponds to the real experiments, 
		and right-hand side 
		represents the toy models discussed in Sec.~\ref{sec:sec2}.
		The dotted arrows show the procedures repeated for iterations.
		The alphabets shown in the boxes correspond to those in Fig.~\ref{fig:flowchart_pic}.
		Numbers in the square brackets are the same as the bullet number for the 
		toy-MC procedures in Sec.~\ref{sec:sec2}.
		}
	\label{fig:flowchart}
\end{figure*}

\begin{figure*}[htbp]
	\begin{center}
	\includegraphics[width=160mm]{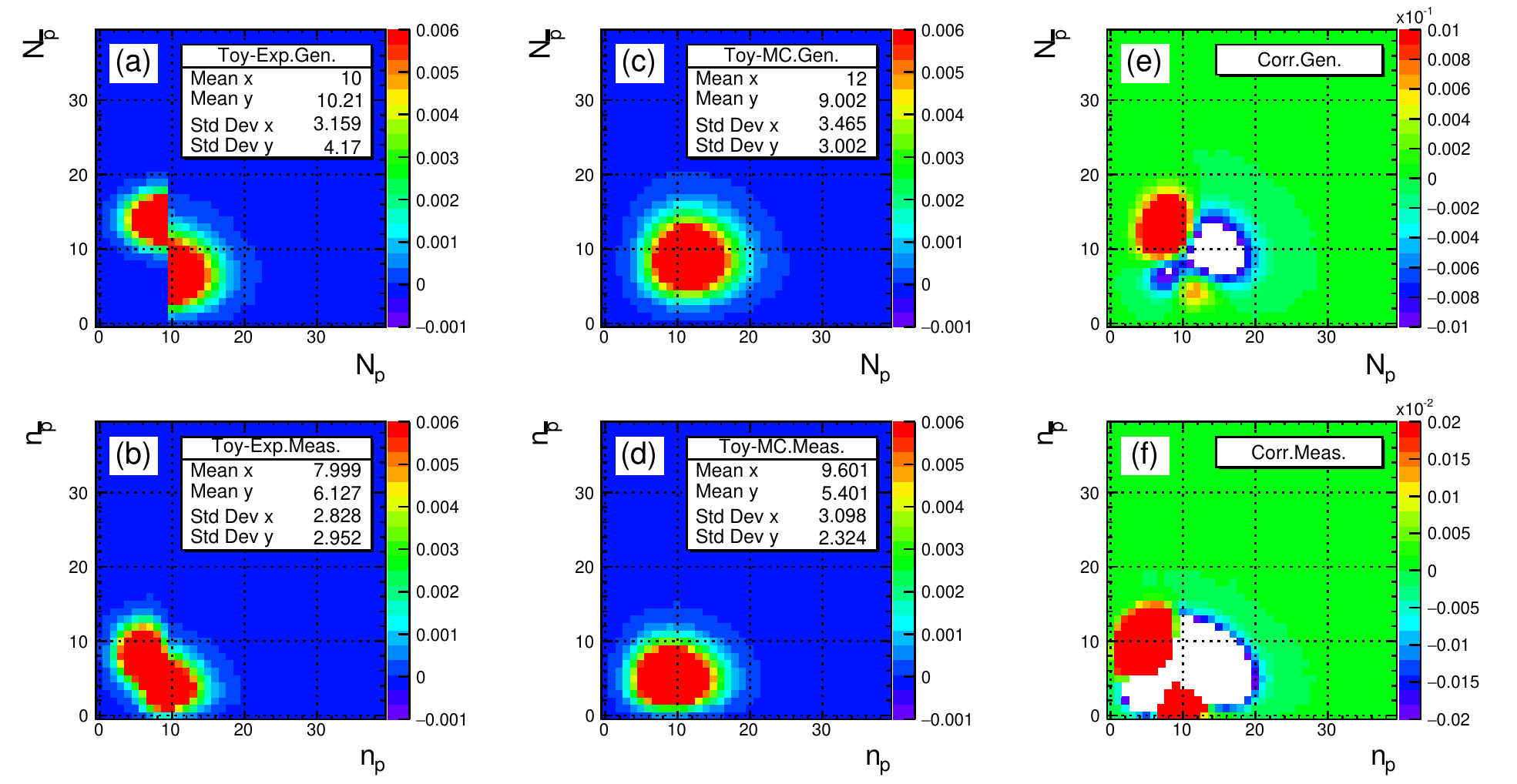}
	\end{center}
	\caption{
		Distributions of particle and antiparticle number
		at various steps in the unfolding procedures:
		(a) generated distribution for toy-experiment,  
		(b) measured distribution for toy-experiment, 
		(c) generated distribution for toy-MC, 
		(d) measured distribution for toy-MC, 
		(e) correction function in the generated coordinates, and
		(f) correction function in the measured coordinates.
		All distributions are normalized.
		Values of mean and standard deviation in x and y axis are shown in the box.
		White-colored bins in panel (b) represent the negative value.
		}
	\label{fig:flowchart_pic}
\end{figure*}

Figure~\ref{fig:flowchart} shows a flowchart of the unfolding procedure.
As shown in the left half, the measured distributions in the real experiments 
are modified by detectors, hence it's distorted from the true distributions. 
Our goal is to obtain this unknown distribution.
The toy models presented in the rest of this section utilize two sets of data.
The first one is the distribution corresponding to experiments, 
which is indicated by "Toy-Experiment" in Fig.~\ref{fig:flowchart}. 
The other one is the virtual (simulation) distribution indicated by "Toy-MC". 
Figure~\ref{fig:flowchart_pic} shows the distributions of  
($N_{p}$,$N_{\bar p}$) or ($n_{p}$,$n_{\bar p}$) at various steps of the unfolding approach, 
corresponding to (a)--(f) indicated in the boxes in Fig.~\ref{fig:flowchart}.
Let us explain the unfolding procedures based on Figs.~\ref{fig:flowchart} 
and \ref{fig:flowchart_pic}.

First, we generate the toy-experiment true distribution (a)  
with the critical shape. 
The Poisson distributions are generated for $P(N_{p})$, 
and for $P(N_{{\bar p}})$ with $N_{p}>10$, 
while the Gauss distributions are generated for $P(N_{{\bar p}})$ with $N_{p}<10$. 
The detector filter with $\varepsilon_{p}=0.8$ and $\varepsilon_{\bar p}=0.6$ 
is applied to get the toy-experiment measured distribution in (b). 

\begin{figure*}[htbp]
	\begin{center}
	\includegraphics[width=130mm]{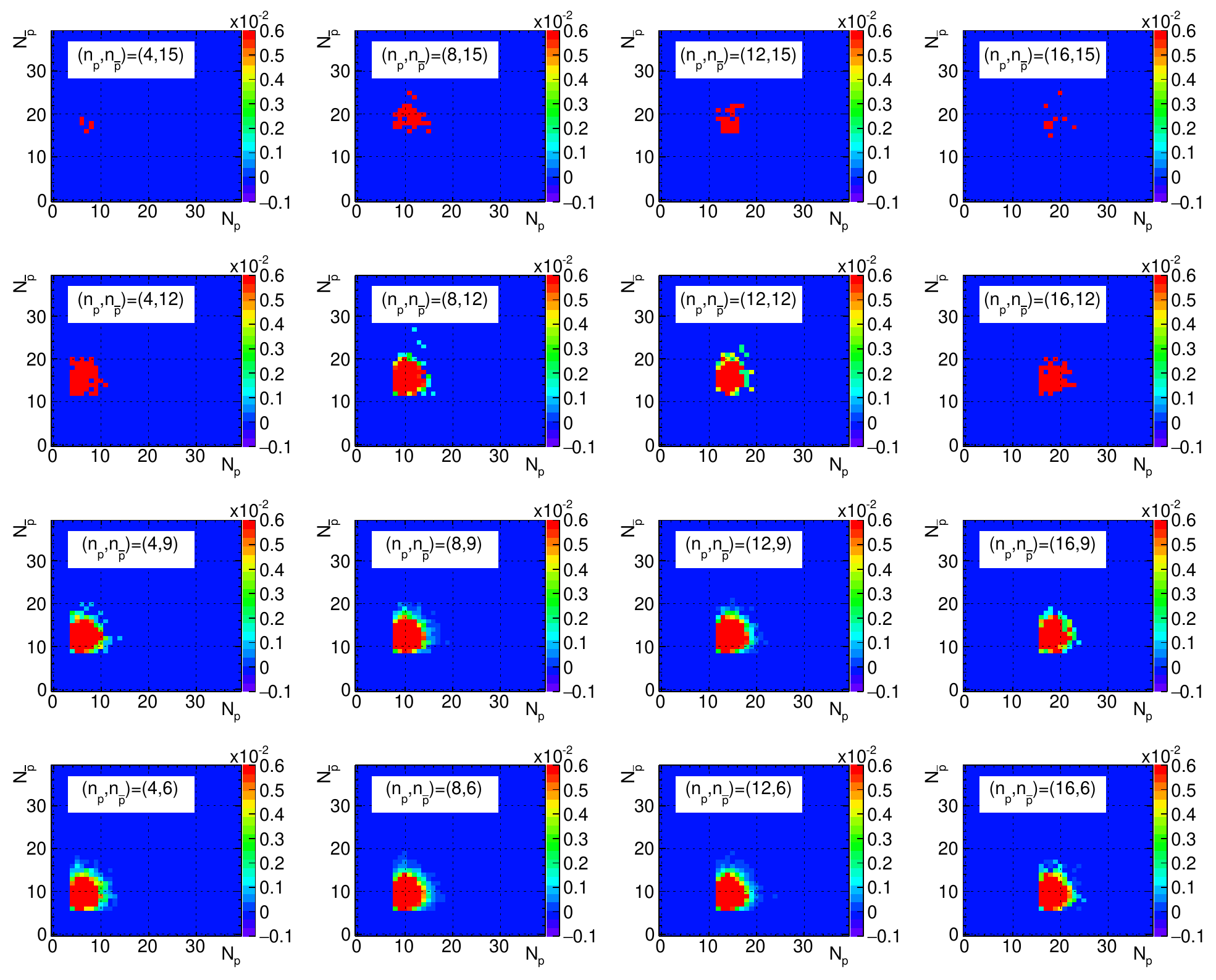}
	\end{center}
	\caption{
		Reversed response matrices, ${\cal R}_{\rm rev}(N_{p},N_{{\bar p}};n_{p},n_{{\bar p}})$, 
		with respect to ($N_{p}$,$N_{{\bar p}}$) for fixed $(n_{p},n_{{\bar p}})$ 
		for the 1st iteration in Sec.~\ref{sec:sec2}. All distributions are normalized to unity. 
		}
	\label{fig:RmRev}
\end{figure*}

The rest of the procedure are iterations to be repeated many times, 
which are shown by the dotted arrows in Fig.~\ref{fig:flowchart}.
Let us explain these procedures as follows. 
The bullet number corresponds to numbers with square brackets in Fig.~\ref{fig:flowchart}.
\begin{description}
	\item[1] Generate a toy-MC distribution, according to a Poissonian 
		with the mean values being 12 and 9 for particles and antiparticles, respectively.
	\item[2] The detector filter is applied to (c) to get toy-MC measured distributions as shown in (d).
	\item[3] During the MC process from \textbf{1} to \textbf{2}, we compute the reversed response matrices, 
		${\cal R}_{\rm rev}$, numerically without any inversion procedure.
		Some examples of the response matrices are shown in Fig.~\ref{fig:RmRev}.
		Each panel shows the probability distributions of ($N_{p}$,$N_{\bar p}$) 
		for the fixed ($n_{p}$,$n_{\bar p}$), which can be directly computed in the MC process.
	\item[4] The correction function is determined by subtracting (d) from (b). See (f).
		It represents the difference between toy-experiment and 
		toy-MC in the measured coordinates. 
	\item[5] We then multiply ${\cal R}_{\rm rev}$ to (f) to get (e) the correction 
		functions in the generated coordinates. 
		It should be noted that 
		smoothing is applied to the correction functions to make the resulting distribution smooth. 
		Further, we multiply a scaling factor $\alpha_{\rm sc}<1$ to (e) before it is added to (c) 
		in order to avoid possible negative content of bins.
		Details on those parameters will be discussed in the following subsections. 
	\item[6] By adding (e) to (c), the toy-MC distribution is modified to be closer to (a).
	\item[7] We repeat (1)--(6) until cumulants of the toy-MC net-particle distribution converge.
		Reversed response matrices are updated for each iteration. 
\end{description}

\begin{figure*}[htbp]
	\begin{center}
	\includegraphics[width=180mm]{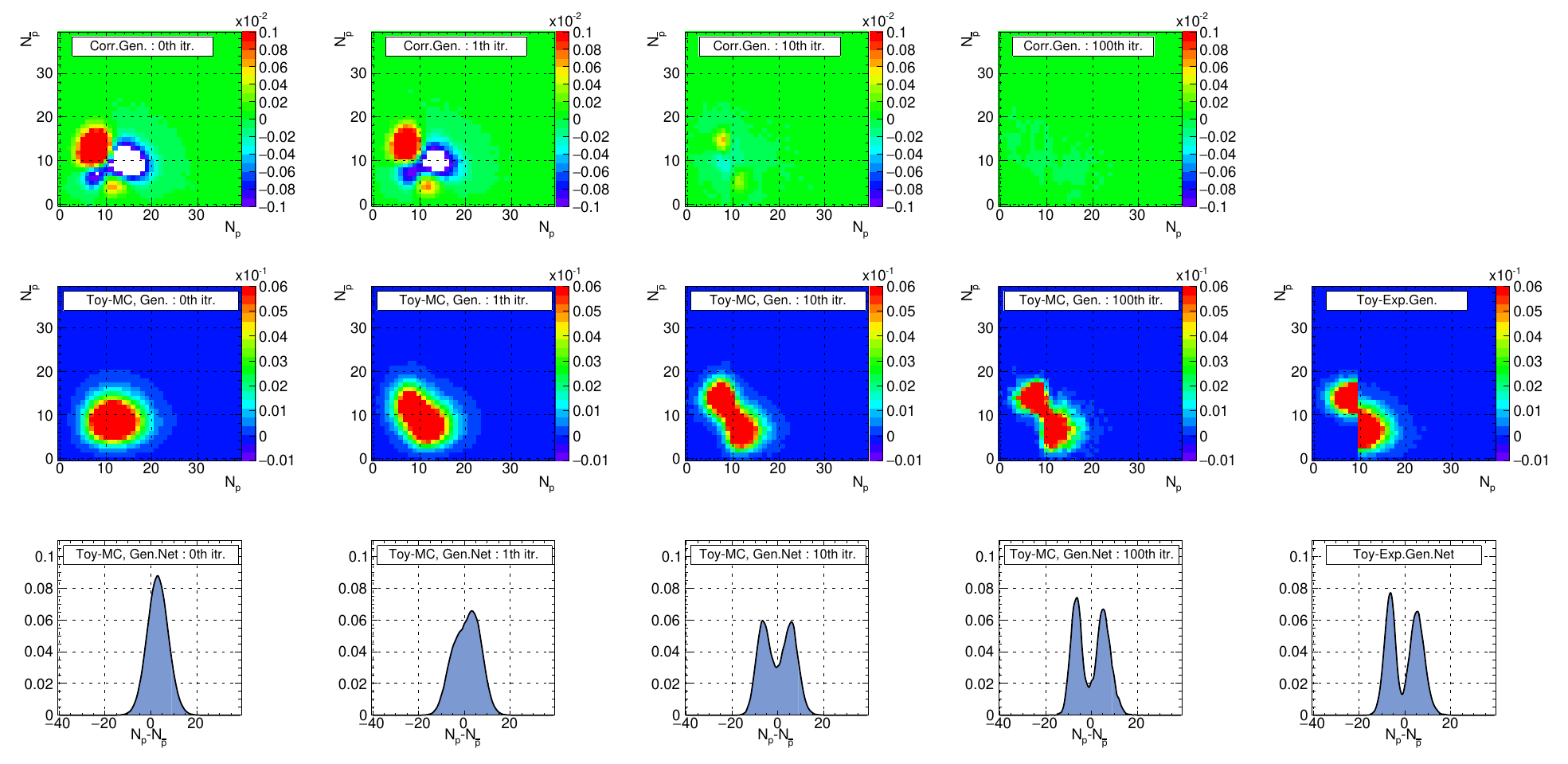}
	\end{center}
	\caption{ 
	(Top) Correction functions in the generated coordinates. 
	White-colored bins represent the large negative value outside the z-axis range.
	(Middle) Toy-MC distributions in the generated coordinates. 
	(Bottom) Toy-MC net-particle distributions in the generated coordinates.
	The 1st to 4th row from left to right show distributions 
	at the 0th (initial condition), 1st, 10th and 100th iteration.
	The most right panels show distributions for the toy-experiment sample.
	}
	\label{fig:itr_change}
\end{figure*}

Figure~\ref{fig:itr_change} shows three kinds of distributions from top to bottom. 
The most right panels in the middle and bottom rows show the toy-experiment and 
its net-particle distributions in the generated coordinates, 
and the 1st to 4th panels from left-hand side 
represent the toy-MC distributions at 0th (initial condition), 1st, 10th and 100th iterations. 
The top row in Fig.~\ref{fig:itr_change} shows the correction functions
in the generated coordinates.
The correction function is seen to flatten with increasing iterations, which indicates that the toy-MC
distribution approaches the toy-experiment distribution 
in the generated coordinates. 
The middle row shows the toy-MC distribution in the generated coordinates. 
The critical shape of the toy-experiment distribution 
is found to be successfully reconstructed starting from the Poisson distribution in toy-MC samples. 
The bottom row shows the toy-MC net-particle distribution.
The two-peak structure in toy-experiment samples is 
reproduced in the toy-MC distribution. 

\begin{figure*}[htbp]
	\begin{center}
	\includegraphics[width=170mm]{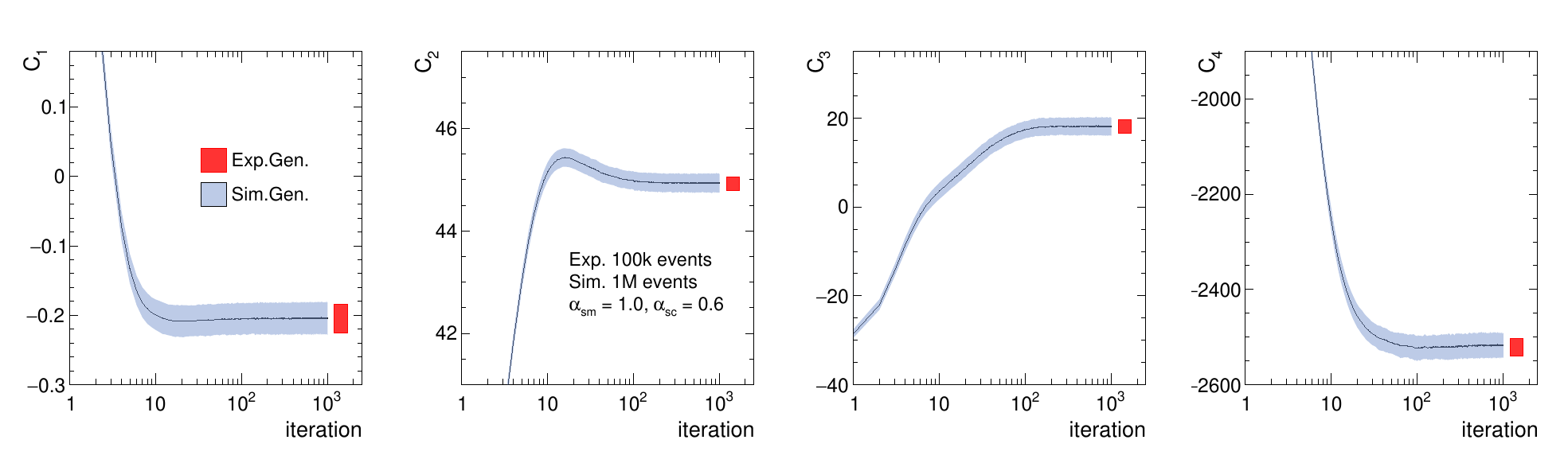}
	\end{center}
	\caption{
		Cumulants up to the 4th-order as a function of iteration. 
		Solid lines and bands show the averaged values and $\pm1\sigma$ 
		for 100 independent trials.
		The box drawn at $x\approx1500$ is the true values of cumulants 
		with $\pm1\sigma$ for the toy-experiment sample in the generated coordinates.
		}
	\label{fig:fig3}
\end{figure*}

Figure~\ref{fig:fig3} shows cumulants up to the fourth-order 
of the toy-MC net-particle distribution in the generated coordinates  
as a function of iteration. 
To see the validity of the unfolding approach, 100 independent samples are generated for 
both toy-experiment and toy-MC samples. 
The averaged values of cumulants are shown in black solid lines, 
and the bands show the statistical uncertainties.
Red boxes show the cumulants of the toy-experiment 
distribution in the generated coordinates.
It is found that the cumulants become flat with increasing iterations 
and consistent with those of the toy-experiment distribution within statistical uncertainties, 
which indicates that our unfolding approach works well.
It is notable that somewhat larger statistical uncertainties for the toy-MC results 
compared to those of the toy-experiment is due to the efficiency loss in the detector filter.
	We note that any efficiency assumptions can be properly taken into account 
	in the unfolding approach. The most important thing is whether or not the detector filters 
	are close enough between toy-experiment and toy-MC samples.
	If this is the case, the unfolding approach works as evident from Fig.~\ref{fig:flowchart} .

\subsection{Smoothing of response matrices}
\begin{figure*}[htbp]
	\begin{center}
	\includegraphics[width=180mm]{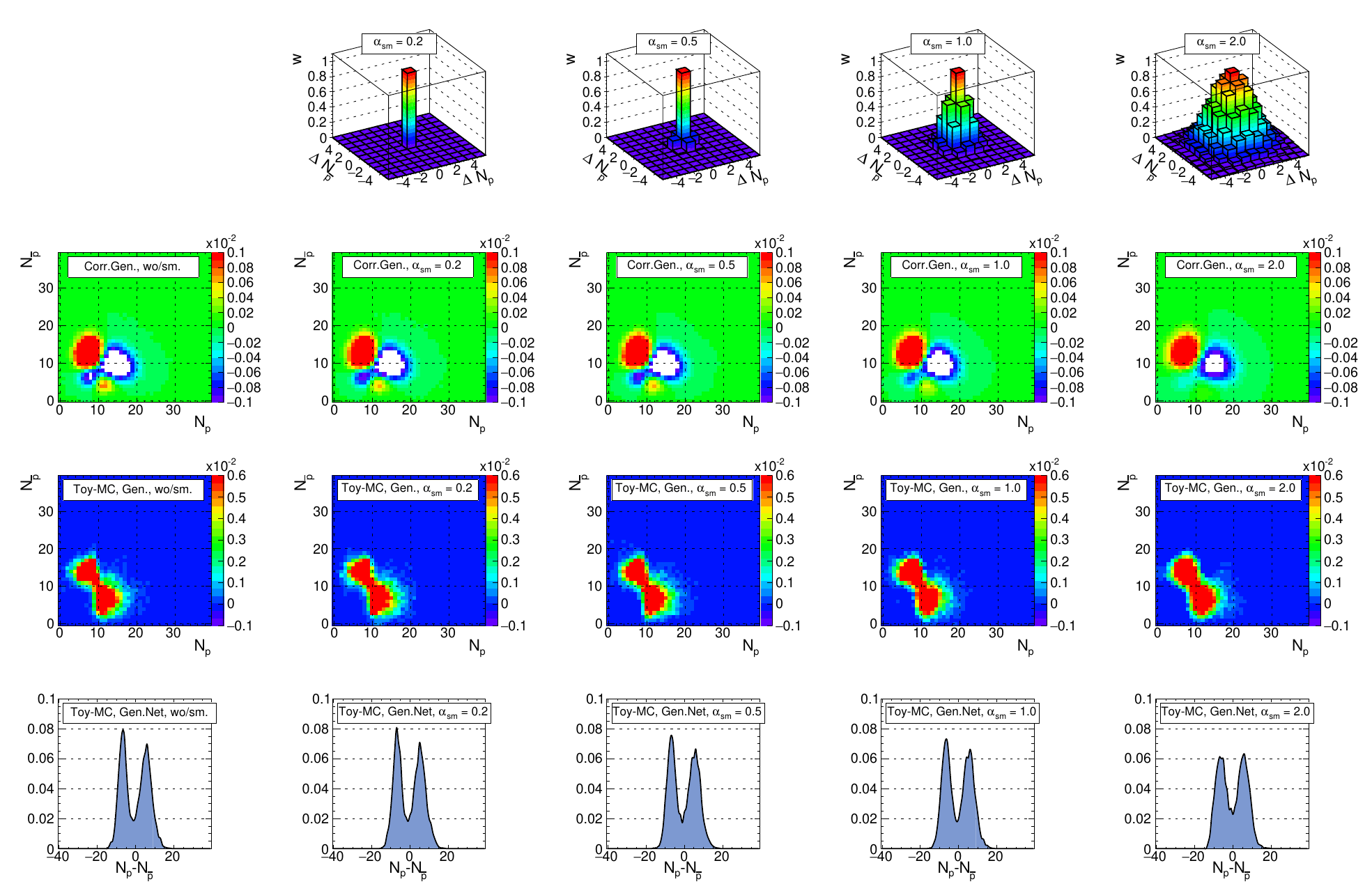}
	\end{center}
	\caption{
		(1st column) Weight function for different smoothing parameter $\alpha_{\rm sm}$. 
		The distribution is normalized so that the value at $(N_{p},N_{{\bar p}})=(0,0)$ becomes unity. 
		(2nd column) Correction functions in the generated coordinates, $P'_{\rm cf}(N_{p},N_{\bar p})$, 
		for different smoothing parameters.
		White-colored bins represent large negative values outside the z-axis range.
		(3rd column) Particle number distributions in the generated coordinate for toy-MC samples 
		at the $80$th iteration.
		(4th column) Net-particle distributions in the generated coordinates for toy-MC samples 
		at the $80$th iteration. 
		The left-most panels show the distributions without smoothing ($\alpha_{\rm sm}=0$).
		}
	\label{fig:sm}
\end{figure*}

\begin{figure*}[htbp]
	\begin{center}
	\includegraphics[width=170mm]{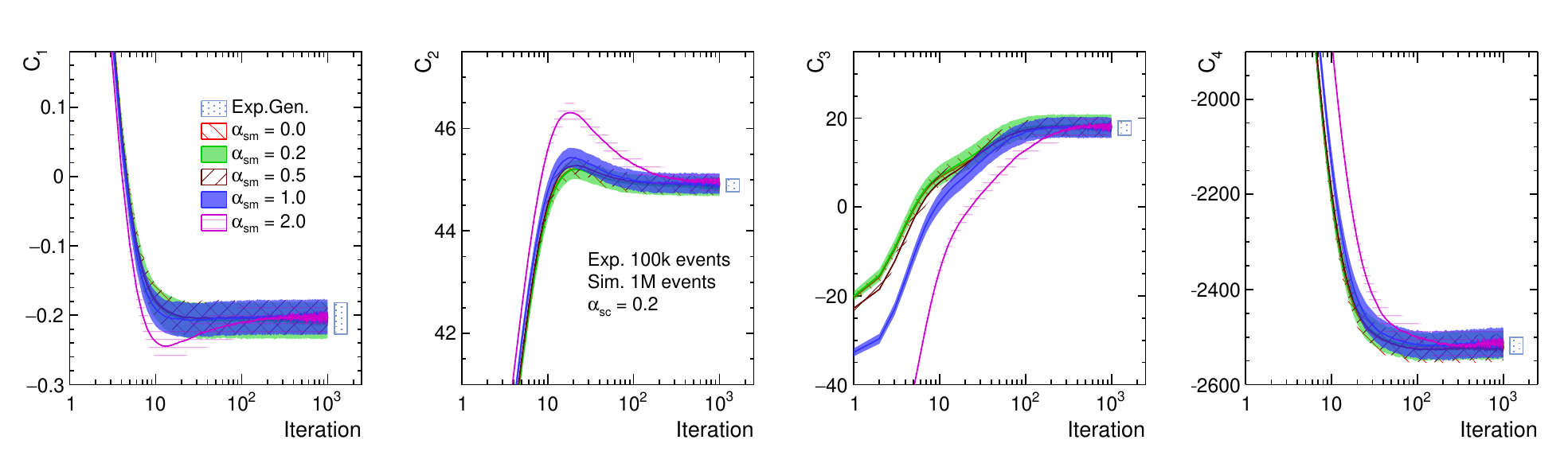}
	\end{center}
	\caption{
		Influence of smoothing on cumulants up to the 4th-order as a function of iteration. 
		Solid lines and bands show the averaged value and $\pm1\sigma$ of the statistical uncertainties 
		for 100 independent trials.
		Different band shadings represent results from different smoothing parameters $\alpha_{\rm sm}$.
		The box drawn at $x\approx1500$ is the true value with $\pm1\sigma$ 
		for the toy-experiment samples in the generated coordinates.
		}
	\label{fig:smitr}
\end{figure*}

The smoothing parameter is used to ensure resulting distributions are smooth.
Since our focus is on the shape of the distribution itself, it is not useful 
to smooth the toy-MC distribution directly. Instead, we try smoothing the correction function:
\begin{eqnarray}
	P'_{\rm cf}(N_{p},N_{\bar p}) &=& 
	 \cfrac {\sum_{x=N_{p}-5}^{N_{p}+5}\sum_{y=N_{\bar p}-5}^{N_{\bar p}+5}w(r,\alpha_{\rm sm})P_{\rm cf}(N_{p},N_{\bar p})}{\sum_{x=N_{p}-5}^{N_{p}+5}\sum_{y=N_{\bar p}-5}^{N_{\bar p}+5}w(r,\alpha_{\rm sm})}, 
\end{eqnarray}
where $P_{\rm cf}(N_{p},N_{\bar p})$ and $P'_{\rm cf}(N_{p},N_{\bar p})$ are correction functions 
at the generated coordinate before and after the smoothing, and $w(r,\alpha_{\rm sm})$ is 
the weight function defined as two-dimensional Gaussian centered at ($N_{p}$,$N_{\bar p}$). 
The $w(r,\alpha_{\rm sm})$ is characterized by $r=\sqrt{(x-N_{p})^{2}+(y-N_{\bar p})^{2}}$ 
and the standard deviation $\alpha_{\rm sm}$. For the former, $x$ and $y$ are defined in a $11\times 11$ matrix region with 
$\pm$5 neighboring bins from the central bin ($N_{p}$,$N_{\bar p}$) 
which corresponds to (0,0) in the top row in Fig.~\ref{fig:sm}. 
The correction factor is determined by averaging the surrounding bins, 
which leads to a smooth distribution after the correction.
A larger value of $\alpha_{\rm sm}$ gives more flattened correction functions.
Figure~\ref{fig:sm} shows the smoothing functions, correction functions, 
toy-MC distributions, and toy-MC net-particle distributions, 
after substantial iterations for $\alpha_{\rm sm}=0$, $0.2$, $0.5$, $1.0$ and $2.0$. 
The correction functions are more smeared with larger smoothing parameters, 
leading to smoother toy-MC distributions. 
This is also visible in toy-MC net-particle distributions, 
where the dip structure around $N_{p}-N_{\bar p}\approx0$ 
is found to be smeared and higher relative to the two peaks with larger value of $\alpha_{\rm sm}$.
Figure~\ref{fig:smitr} shows the cumulants up to the 4th order as a function of iteration 
with different smoothing parameters.
Although the convergence behavior seems to change with different smoothing parameters, the cumulant values are consistent 
for all cases after 1000 iterations within statistical uncertainty.
This indicates that the smoothing process would not affect the final results of cumulants, 
but we propose to try several smoothing parameters to test the stability of the 
final results.

\subsection{Scaling of response matrices}
\begin{figure*}[htbp]
	\begin{center}
	\includegraphics[width=170mm]{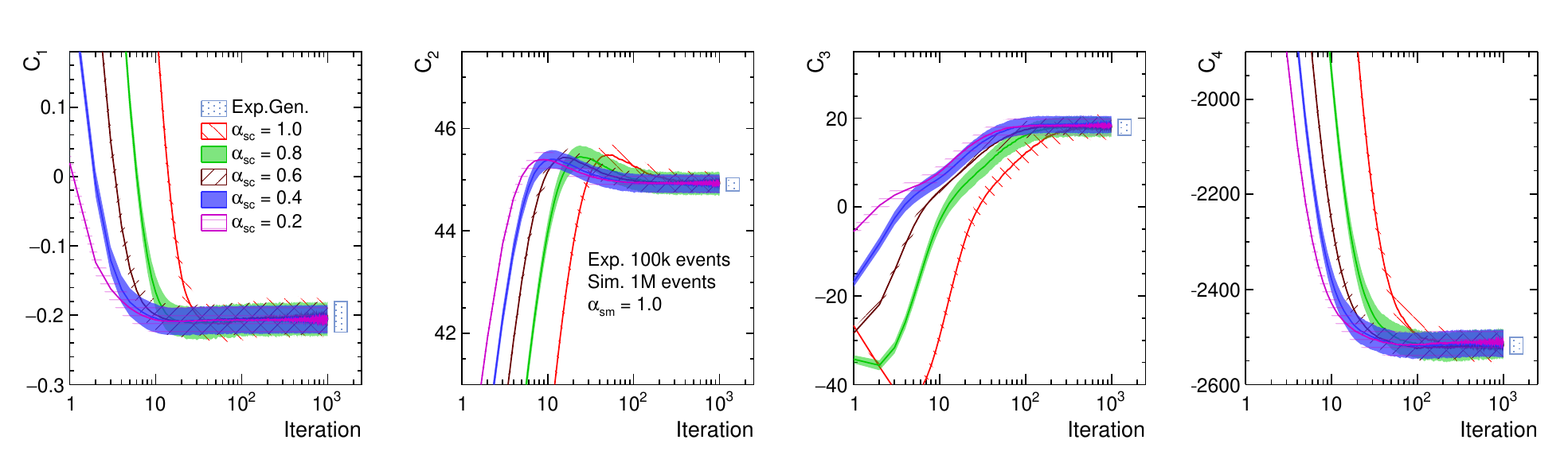}
	\end{center}
	\caption{
		Influence of scaling on cumulants up to the 4th-order as a function of iteration. 
		Solid lines and bands show the averaged value and $\pm1\sigma$ of the statistical uncertainties 
		for 100 independent trials.
		Different band shadings present results from different scaling parameters $\alpha_{\rm sc}$.
		The box drawn at $x\approx1500$ is the true value with $\pm1\sigma$ 
		for the toy-experiment samples in the generated coordinates.
		}
	\label{fig:sc}
\end{figure*}

Another parameter for scaling is introduced to avoid negative content in the 
toy-MC distribution after correction functions are applied.
As one can see from Fig.~\ref{fig:flowchart_pic}-(e) and (f), the correction functions have 
some negative values shown in blue or white-colored bins. Negative values can thus 
appear in the resulting toy-MC distributions in Fig.~\ref{fig:flowchart_pic}-(c). 
The negative values in the toy-MC distributions affect the MC process 
in the next iteration. 
The detector filter cannot be applied to bins with a negative number 
of events. We therefore set bins to be artificially zero in these cases.
In order to avoid such special treatment as much as possible, the correction functions are scaled down by the 
parameter $\alpha_{\rm sc}<1.0$.
Figure~\ref{fig:sc} shows the cumulants up to the 4th order as a function of iteration 
with different scaling parameters. It is found that the scaling parameter controls the 
convergence speed, but the cumulant values are consistent within uncertainties 
after substantial iterations.
	Therefore, the scaling process does not greatly affect the final results. 
	A large scaling value is better if one wants to save calculation cost. 
	However, even then we advice to check the consistency of the final results with several 
	scaling parameters.

\subsection{Application to bimodal distribution}
\begin{figure*}[htbp]
	\begin{center}
	\includegraphics[width=90mm]{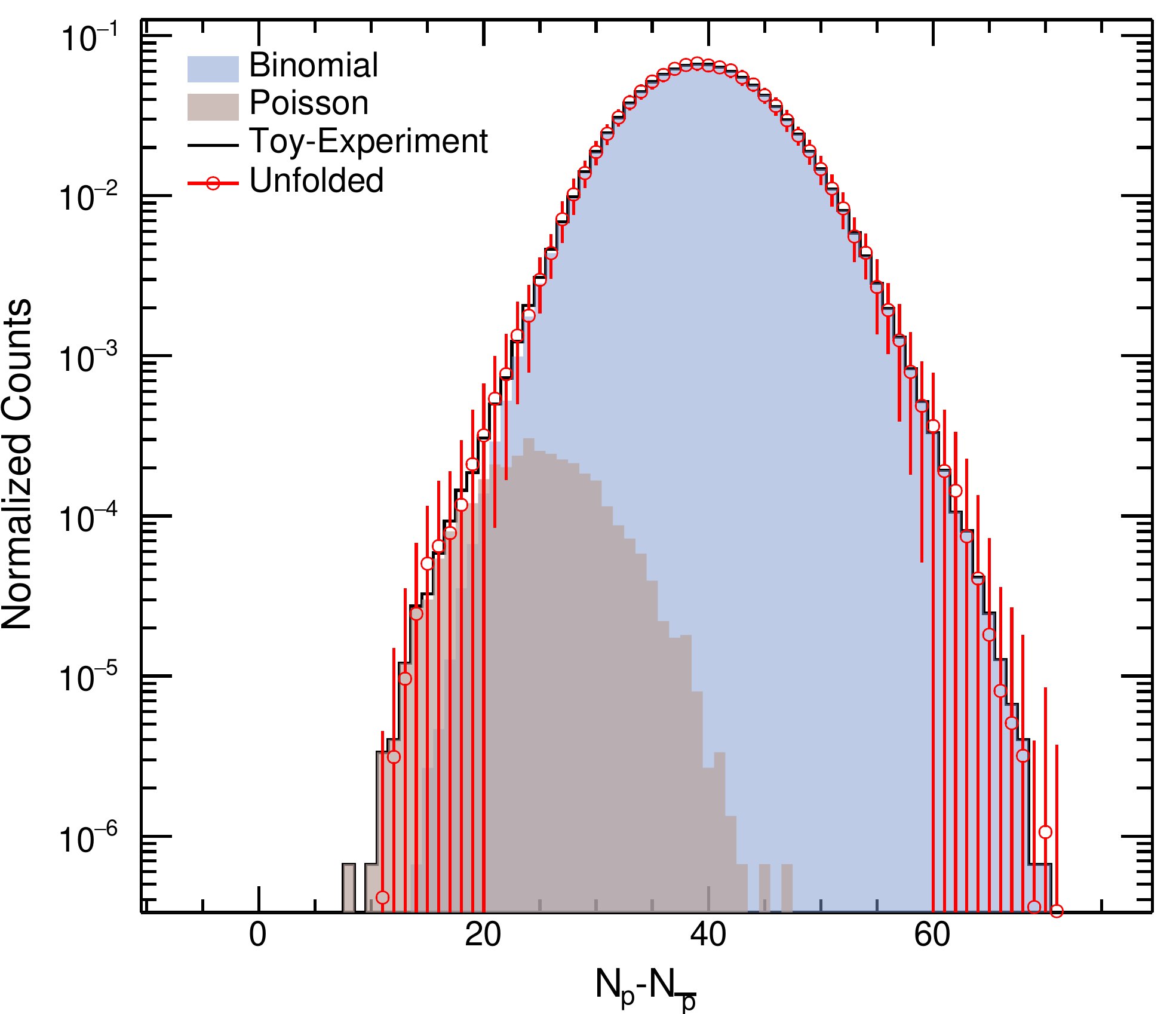}
	\end{center}
	\caption{
		Net-particle distributions for toy-experiment (black) and unfolded toy-MC (red) samples.
		The binomial and Poisson distributions which form the toy-experiment distribution 
		are shown in shaded histograms.
		}
	\label{fig:bimodal}
\end{figure*}

In order to check the sensitivity of the unfolding approach in a more realistic case, 
we applied another model assuming a bimodal distribution 
for the toy-experiment samples. 
According to Ref.~\cite{Bzdak:2018uhv}, the large value of $C_{4}/C_{2}$ observed by the 
STAR experiment in Au+Au central collisions at $\sqrt{s_{\rm NN}}=7.7$~GeV 
can be described by the superposition of binomial and Poisson distributions: 
\begin{eqnarray}
	P_{ab}(N) &=& (1-w)P_{a}(N) + wP_{b}(N), \\ 
	P_{a}(N) &=& \cfrac{B!}{N!(B-N)!}\varepsilon^{N}(1-\varepsilon)^{B-N},\;\; 
	P_{b}(N) = \cfrac{\lambda^{N}e^{-\lambda}}{N!}
\end{eqnarray}
where $P_{a}$ and $P_{b}$ are the binomial and Poisson distributions, 
with $w=0.0033$, $B=350$, $\varepsilon=0.114$ and $\lambda=25.3$.
We generate the toy-experiment distribution according to $P_{ab}(N)$ for particles, 
and another Poisson distribution for antiparticles with the mean value 
being $0.3$ taken from Ref.~\cite{Adam:2020unf}.
The same detector filter described in Sec.~\ref{sec:critical} is used. 
The scaling and smoothing parameters are chosen to be $\alpha_{\rm sc}=0.6$ and $\alpha_{\rm sm}=0$.
For the toy-experiment samples 150~k events are generated, while 1.5~M events 
are generated for the toy-MC distribution. The 100 independent samples are generated to check statistical uncertainties.
Results of the cumulants up to the 4th order are shown in Tab.~\ref{tab:bimodal} for toy-experiment and unfolded toy-MC 
distribution after 1000 iterations.
We checked that the cumulants are converged reasonably after the iterations.
Cumulant values for the unfolded toy-MC distribution are found to be consistent with those from 
the toy-experiment distribution.
\begin{table}[htb]
  \begin{tabular}{ccc} \hline
	  Cumulant & Toy-Experiment$\pm$stat.err & Toy-MC (Unfolded)$\pm$stat.err \\ \hline
$C_{1}$ & 39.554 $\pm$ 0.014  & 39.555 $\pm$ 0.017 \\ \hline 
$C_{2}$ & 36.33$\pm$0.15 & 36.32$\pm$0.18 \\ \hline 
$C_{3}$ & 18.4$\pm$1.6  & 18.3$\pm$2.2  \\ \hline 
$C_{4}$ & 118 $\pm$ 22 & 112 $\pm$ 33   \\ \hline
  \end{tabular}
	\caption{Cumulants up to the 4th-order for the bimodal toy-experiment 
	distribution and toy-MC distribution after 1000 iterations (see Fig.~\ref{fig:bimodal}).}
	\label{tab:bimodal}
\end{table}
Corresponding net-particle distributions for toy-experiment and unfolded toy-MC samples 
are shown in Fig.~\ref{fig:bimodal}.
The binomial and Poisson distributions forming the toy-experiment distributions are also shown.
The data points for the unfolded distribution were calculated by averaging the results from 100 
independent samples, while the statistical uncertainties are for one sample with 150k events.
It is demonstrated that the toy-MC distribution is successfully unfolded 
having a small bump at the low side of the distribution with 
15M and 150M effective events for toy-experiment and toy-MC samples, respectively.
Please note the data points would be expected to fluctuate as shown by the  
large statistical uncertainties in Fig.~\ref{fig:bimodal} 
in the case of only 150k events for the toy-experiment sample. 

\section{Volume fluctuation convoluted unfolding\label{sec:sec3}}
In this section, we expand the particle number unfolding discussed in Sec.~\ref{sec:sec2} 
to deal with volume fluctuations.
The following subsections first define the volume fluctuations then discuss 
the methodology.
\subsection{Volume fluctuation}
The initial volume in heavy-ion collisions is characterized by 
the impact parameter $b$, which is defined as the distance between the center of two nuclei.
We also consider the number of participant nucleons, 
spectator nucleons and binary collisions based on the Glauber model. 
These variables are not accessible directly in experiments. 
Thus, the particle production model is utilized to 
produce the final state multiplicity from sources, $N_{\rm source}$, 
which is usually expressed in terms of participant nucleons and binary collisions, or 
a power function of the former. 
The resulting multiplicity distribution is then compared to 
experimentally measured multiplicity distribution to define the centrality.
In this case, one can easily imagine that the value of $N_{\rm source}$ 
fluctuates even with fixed multiplicity. This is called volume fluctuation.
Assuming that particles are produced from the independent source of $N_{\rm source}$ in a fixed volume, 
the true cumulants of particle distributions are expressed 
by superposition of cumulants for each source
\begin{equation}
	C_{r}(\Delta N) = \sum_{N_{\rm source}}\kappa_{r}(\Delta m)
\end{equation}
with
\begin{equation}
	\Delta N = N_{p} - N_{{\bar p}}, \;\; \Delta m = m_{p} - m_{{\bar p}},
\end{equation}
where $m_{p}$ and $m_{{\bar p}}$ are particles and antiparticles produced 
per participant nucleon.
In this situation, the cumulants can be analytically corrected for volume fluctuations
~\cite{Skokov:2012ds,Braun-Munzinger:2016yjz}.
On the other hand, the particle distribution cannot be corrected in this way. 
In the next subsection we explain how to implement volume fluctuations 
into the particle number unfolding.

\subsection{Methodology}
\begin{figure*}[htbp]
	\begin{center}
	\includegraphics[width=160mm]{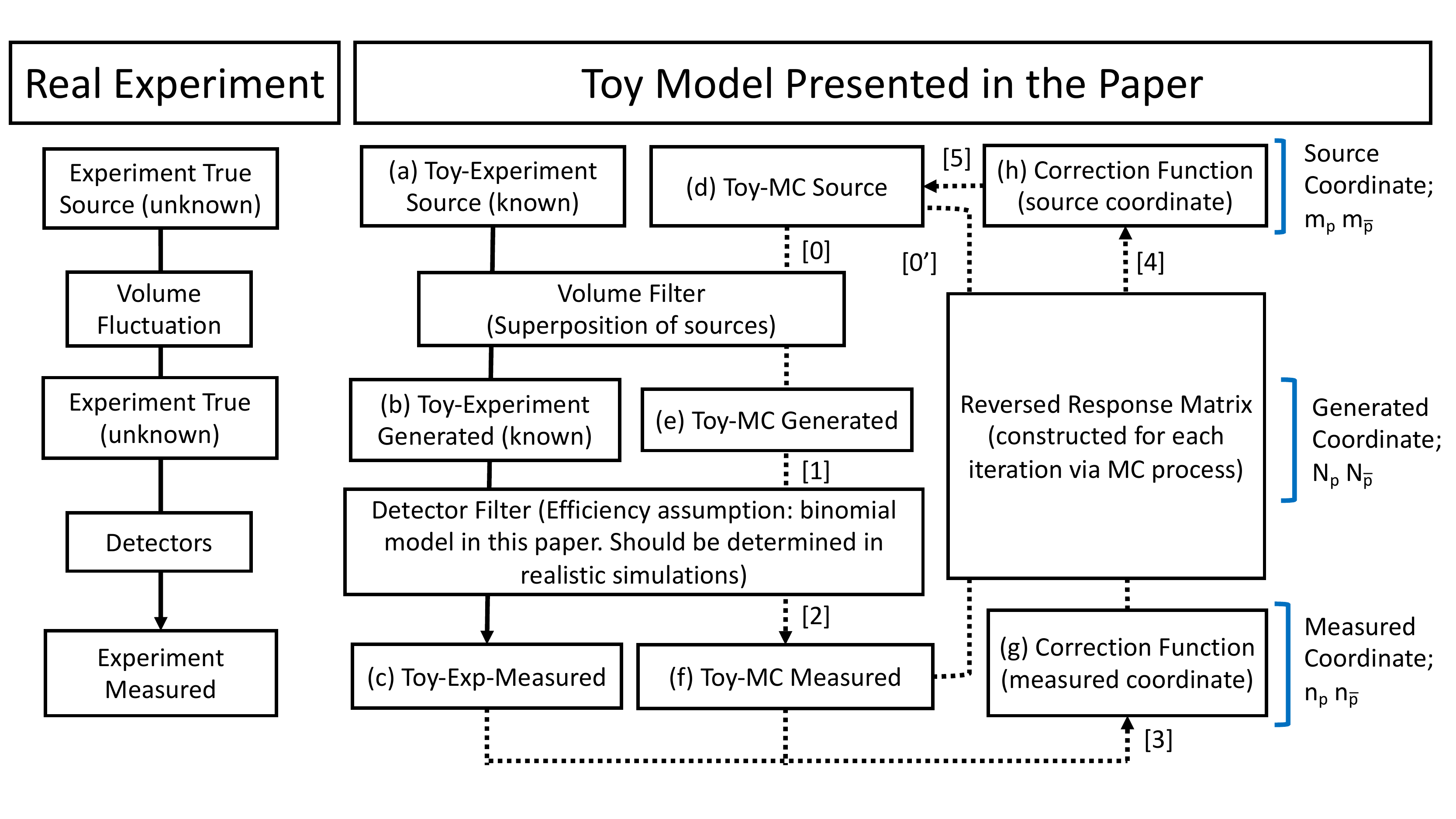}
	\end{center}
	\caption{
		Flowcharts in the particle number unfolding convoluting volume fluctuations. 
		The dotted arrows show the procedures repeated for iterations.
		The alphabets shown in the boxes correspond to those in Fig.~\ref{fig:VF_proc2}.
		Numbers in the square brackets are the same as the bullet numbers for 
		toy-MC procedures in Sec.~\ref{sec:sec3}.
		}
	\label{fig:flowchart_vf}
\end{figure*}

\begin{figure*}[htbp]
	\begin{center}
	\includegraphics[width=130mm]{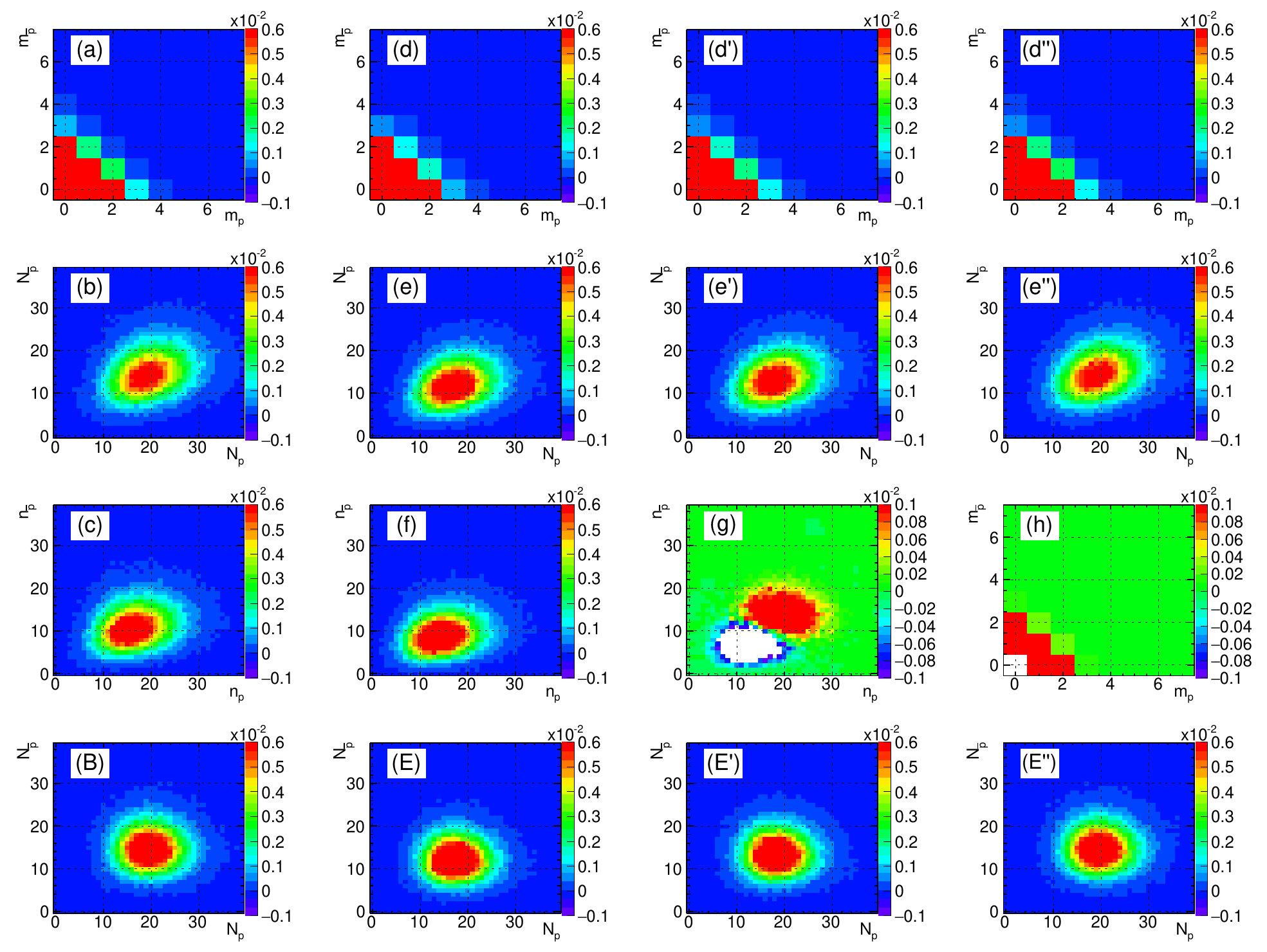}
	\end{center}
	\caption{
		Particle and antiparticle number distributions of toy-experiment samples for 
		(a) each source 
		(b) superposition for fluctuating $N_{\rm source}$
		(B) superposition for fixed $N_{\rm source}$ 
		(c) superposition for fluctuating $N_{\rm source}$ with the detector filter.
		The same plots are shown for toy-MC samples in panels (d)--(f) and (E).
		Panel (g) shows the correction function in the measured coordinates. 
		Panel (h) shows the correction function in the source coordinates.
		Panels (d'), (e') and (E') are are the distributions after the 1st iteration.
		Panels (d''), (e'') and (E'') are are the distributions after the 2nd iteration.
		White-colored bins in panels (g) and (h) represent large negative values outside the z-axis range.
		}
	\label{fig:VF_proc2}
\end{figure*}

Let us start from the toy-experiment distribution in the source coordinates. 
Since the approach of particle number unfolding was found to work well 
for both the extreme and the realistic cases in Sec.~\ref{sec:sec2}, 
we focus on the simple negative binomial distributions 
for source distributions~\cite{Ansorge:1988kn}. 
The particle distributions per source for toy-experiment samples are generated with 100k 
events based on the negative binomial distribution:
\begin{equation}
	P_{\mu,k}(m) = \cfrac{\Gamma(m+k)}{\Gamma(m+1)\Gamma(k)}\cdot\cfrac{(\mu/k)^{m}}{(1+\mu/k)^{m+k}},  
\end{equation}
where $m$ is the particle number per source.
Parameters are chosen to be ($\mu_{p}$,$k_{p}$)$=$(0.2,5.0) and ($\mu_{\bar p}$,$k_{\bar p}$)$=$(0.15,3.0) for 
particles and antiparticle, respectively.
The particle distribution per source for the toy-experiment sample is shown in Fig.~\ref{fig:VF_proc2}-(a).
The particle distributions are generated $N_{\rm source}$ times, which are superimposed to produce the 
toy-experiment distribution in the generated coordinates.
We tried two ways to superimpose the particle distributions. 
In one case the value of $N_{\rm source}$ is fixed for all events, 
while in another $N_{\rm source}$ is defined as the distribution of participant nucleons from 
the Glauber model for Au+Au collisions with the impact parameter $\sim9$~fm. 
This process of superposition will be referred to as the "volume filter" at the rest of this paper. 
Both cases are shown in Fig.~\ref{fig:VF_proc2}-(b) and (B), respectively. 
The positive correlation of $N_{p}$ and $N_{\bar p}$ in Fig.~\ref{fig:VF_proc2}-(b) indicates 
the effect of volume fluctuations.
We apply the detector filter as performed in Sec.~\ref{sec:sec2} to determine the 
toy-experiment distributions in the measured coordinates as shown in Fig.~\ref{fig:VF_proc2}-(c).
The rest procedures are performed with toy-MC distributions, which are explained as follows.
The bullet numbers correspond to those with square brackets in Fig.~\ref{fig:flowchart_vf}. 
Figure \ref{fig:flowchart_vf} depicts a flowchart of the unfolding procedure performed 
with the toy-MC distributions shown in Fig.~\ref{fig:VF_proc2}.
\begin{description}
	\item[0] Generate a toy-MC distribution per source (100k events), according to the negative binomial distribution 
		with parameters randomly fluctuated with $\pm20$\% compared to those in the toy-experiment samples. 
		See Fig.~\ref{fig:VF_proc2}-(d).
	\item[1] The volume filter is applied to (a) to get toy-MC distributions in the generated coordinates. 
		Two kinds of volume filters with and without volume fluctuations 
		are applied as explained for toy-experiment samples above.
		Resulting toy-MC distributions in the generated coordinates are shown in Fig.~\ref{fig:VF_proc2}-(e) and (E).
	\item[2] The detector filter is applied to (e) to get toy-MC measured distributions as shown in Fig.~\ref{fig:VF_proc2}-(f).
	\item[0'] During the MC process from \textbf{0} to \textbf{2}, the reversed response matrices 
		${\cal R}_{\rm rev}({m_{p},m_{{\bar p}};n_{p},n_{{\bar p}}})$ 
		are computed.  The toy-MC distributions in the measured 
		and source coordinates are connected as the following relation: 
		\begin{equation}
			P(m_{p},m_{{\bar p}}) = \sum_{n_{p},n_{{\bar p}}}{\cal R}_{\rm rev}({m_{p},m_{{\bar p}};n_{p},n_{{\bar p}}}){\tilde P}(n_{p},n_{{\bar p}}).
		\end{equation}
		The important point here is that the generated coordinates ($N_{p}$,$N_{\bar p}$) is skipped 
		in the response matrices unlike the unfolding approach discussed in Sec.~\ref{sec:sec2}.
	\item[3] The correction function is determined by subtracting 
		Fig.~\ref{fig:VF_proc2}-(c) from (f), which is shown in Fig.~\ref{fig:VF_proc2}-(g).
		It represents the difference between toy-experiment and 
		toy-MC samples in the measured coordinates. 
	\item[4] We then multiply ${\cal R}_{\rm rev}({m_{p},m_{{\bar p}};n_{p},n_{{\bar p}}})$ 
		to Fig.~\ref{fig:VF_proc2}-(g) to get the correction functions for the 
		source coordinates as shown in Fig.~\ref{fig:VF_proc2}-(h) .
			Smoothing and scaling are carried out 
			based on the correction functions as done in Sec.~\ref{sec:sec2}. 
			Parameters are chosen to be $\alpha_{\rm sm}=0.1$ and $\alpha_{\rm sc}=1.0$.
	\item[5] By adding Fig.~\ref{fig:VF_proc2}-(h) to (d), the toy-MC source distribution 
		is modified to be close to (a). See Fig.~\ref{fig:VF_proc2}-(d').
	\item[Iteration] Repeat \textbf{1}--\textbf{5} until cumulants of the toy-MC net-particle distribution converge.
\end{description}

In this way, the toy-MC source distribution (Fig.~\ref{fig:VF_proc2}-(d)) is modified with iterations.
The toy-MC distribution in the generated coordinates with volume filters (Fig.~\ref{fig:VF_proc2}-(e)) 
is also modified accordingly. 

\begin{figure*}[htbp]
	\begin{center}
	\includegraphics[width=140mm]{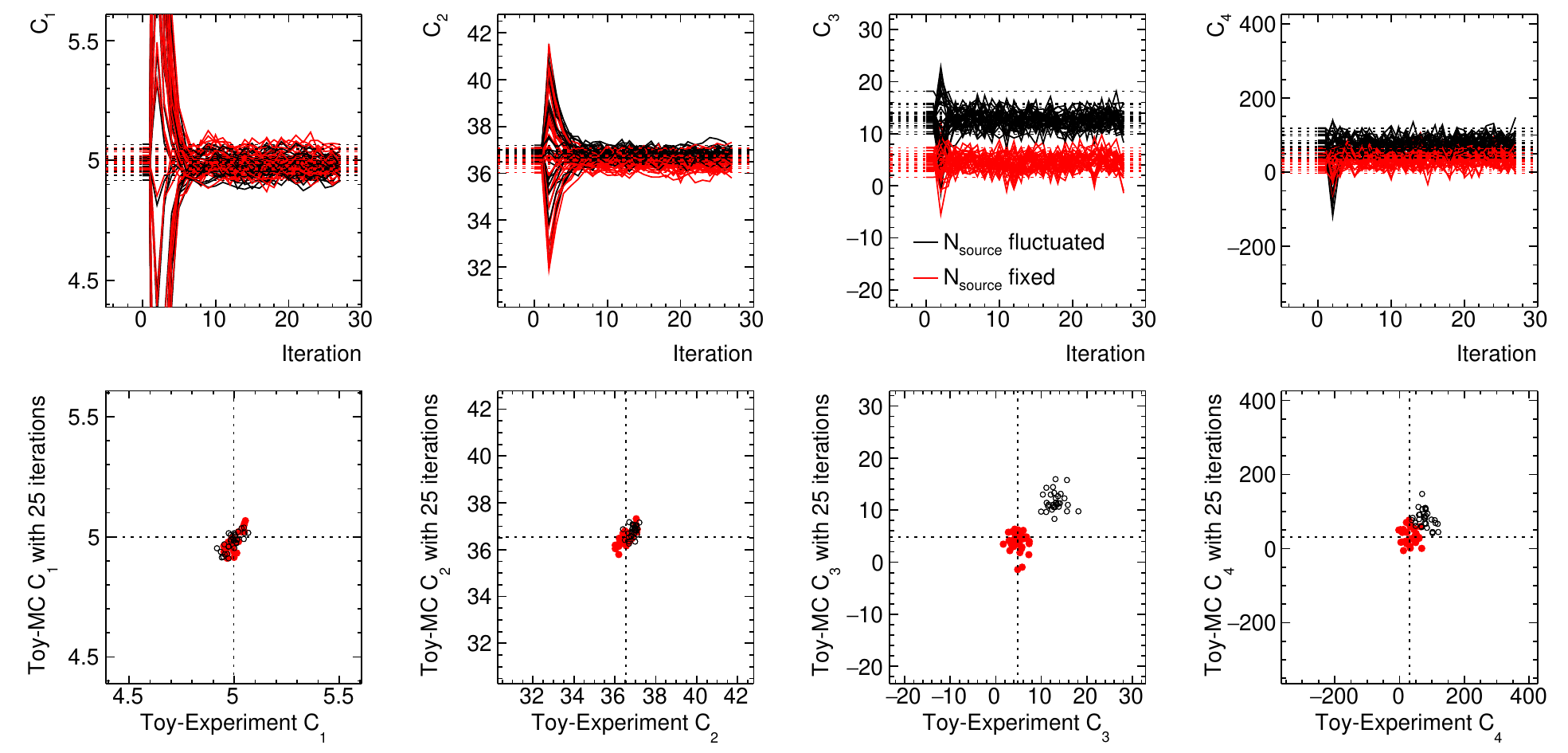}
	\end{center}
	\caption{
		(Top) Lines show the cumulants as a function of iteration. 
		Results from 100 independent trials are superimposed. 
		Dashed-lines show the cumulants of the toy-experiment samples 
		with and without volume fluctuations.  
		(Bottom) Correlation between input and output cumulants.
		Black represents results which include volume fluctuations and 
		red represent results which do not. 
		}
	\label{fig:VF_itr}
\end{figure*}

The top row in Fig.~\ref{fig:VF_itr} shows cumulants up to the fourth order of 
the toy-MC distributions in the generated coordinates as a function of iteration.
The volume fluctuations are convoluted for black lines, while 
no volume fluctuations is taken into account (fixed $N_{\rm source}$) for red lines.
Results from 100 independent samples are plotted to see the statistical fluctuations. 
Dashed-lines from -5 to 30 on x-axis represent the cumulants of toy-experiment 
distributions in the generated coordinates.
Two results are found to be separated for the 3rd and 4th order cumulants, 
which is due to the volume fluctuations.
It is found that the cumulants of toy-MC samples do converge to those 
of toy-experiment samples after 25 iterations. 
Bottom panels show correlation of cumulants for generated coordinates 
between toy-experiment and toy-MC samples after iterations. 
The consistency between x and y axis indicates 
that our unfolding approach convoluting the volume fluctuation does work well.

One final remark is as follows. It was pointed out in Ref.~\cite{Sugiura:2019toh} that the 
independent particle production model would be broken in the UrQMD framework, 
as well as in the real experiment where we expect 
strongly interacting hot and dense matter to form.
From discussions so far, it is obvious that one can implement any volume fluctuations 
in the volume filter. 
	More importantly, one needs to check if the assumed volume fluctuations 
	are reasonable by model simulations. 
	As long as the assumption is correct, our unfolding approach can deal with 
	any kind of well-defined volume fluctuation to 
	reconstruct the true particle number distributions.


\section{Summary}
We presented the particle number unfolding methodology for 
the measurement of higher-order cumulants of net-particle distributions.
The unfolding approach is applied to both an extreme case with very strong 
critical shape and a more realistic case with a small/weak 2nd component. 
The latter shows that our approach can successfully reconstruct 
the bimodal distribution expected in Au+Au collisions 
at $\sqrt{s_{\rm NN}}=7.7$~GeV~\cite{Bzdak:2018uhv,Adam:2020unf}.

We also demonstrated convolution of the volume fluctuations 
through this approach by considering the particle production from independent sources.
Our method should be useful to reconstruct particle number distribution 
in terms of both detector efficiencies and volume fluctuations, 
which could be also extended to reconstruct simultaneously other event-wise quantities like
temperature (mean transverse momentum) fluctuations of the system created in heavy-ion collisions, 
on top of initial volume fluctuation filters as one likes to.

\section{Acknowledgement}
We would like to thank X. Luo, B. Mohanty, A. Pandav, N. Xu and Y. Zhang for fruitful discussions.
We would like to thank Enago (www.enago.jp) for the English language review.
This work was supported by the National Natural Science Foundation of China No 11950410505, 
China Postdoctoral Science Foundation funded project 2018M642878, Ito Science Foundation (2017) and
JSPS KAKENHI Grant No. 25105504 and 19H05598.

\bibliography{main}

\end{document}